\documentclass{elsart}
\usepackage{psfrag,graphicx,float}
\usepackage{amsmath,amssymb}
\journal{Nuclear Physics B}
\newcommand{\be}{\beta}
\newcommand{\de}{\delta}
\newcommand{\ep}{\epsilon}
\newcommand{\vep}{\varepsilon}
\newcommand{\ga}{\gamma}
\newcommand{\ka}{\kappa}
\newcommand{\la}{\lambda}

\newcommand{\si}{\sigma}

\newcommand{\vp}{\varphi}

\newcommand{\La}{\Lambda}
\newcommand{\Si}{\Sigma}
\newcommand{\bchi}{\boldsymbol{\chi}}
\newcommand{\bx}{\mathbf{x}}

\newcommand{\bxi}{\boldsymbol{\xi}}

\newcommand{\bvep}{\boldsymbol{\vep}}
\newcommand{\tC}{\widetilde{C}}
\newcommand{\tK}{\tilde{K}}
\newcommand{\tS}{\tilde{S}}

\newcommand{\tih}{\tilde{h}}

\newcommand{\sse}{\mathsf{e}}
\newcommand{\ssh}{\mathsf{h}}
\newcommand{\ssC}{\mathsf{C}}
\newcommand{\ssH}{\mathsf{H}}
\newcommand{\ssI}{\mathsf{I}}
\newcommand{\ssJ}{\mathsf{J}}
\newcommand{\ssZ}{\mathsf{Z}}
\newcommand{\hJ}{\widehat{J}}
\let\NN\Nset
\let\RR\Rset
\let\ZZ\Zset
\newcommand{\cB}{{\mathcal B}}

\newcommand{\cH}{{\mathcal H}}

\newcommand{\cM}{{\mathcal M}}
\newcommand{\cN}{{\mathcal N}}
\newcommand{\cP}{{\mathcal P}}
\newcommand{\cR}{{\mathcal R}}
\newcommand{\cS}{{\mathcal S}}

\newcommand{\fD}{{\mathfrak D}}

\newcommand{\fK}{{\mathfrak K}}

\newcommand{\fS}{{\mathfrak S}}

\def\Bbe{\overline\beta}

\def\Hs{H_{\mathrm{s}}}
\def\Zs{Z_{\mathrm{s}}}
\newcommand{\pa}{\partial}
\newcommand{\ra}{\to}

\newcommand{\abs}[1]{\left|#1\right|}
\def\ket#1{|#1\rangle}

\let\ds\displaystyle
\let\ni\noindent
\newcommand{\ms}{\mspace{1mu}}
\newcommand{\tr}{\operatorname{tr}}

\newcommand{\card}{\operatorname{card}}

\newcommand{\iu}{{\rm i}}

\begin{document}
\begin{frontmatter}

\title{Haldane--Shastry spin chains of $BC_N$
type}

\author{A. Enciso}, \author{F. Finkel}, \author{A.
  Gonz{\'a}lez-L{\'o}pez}, \author{M.A. Rodr{\'\i}guez}
\address{Depto.~de F{\'\i}sica Te{\'o}rica II, Universidad
Complutense, 28040 Madrid, Spain}
\date{December 9, 2004}

\begin{abstract}
We introduce four types of \hbox{$\mathrm{SU}(2M+1)$} spin chains
which can be regarded as the $BC_N$ versions of the celebrated
Haldane--Shastry chain. These chains depend on two free parameters
and, unlike the original Haldane--Shastry chain, their sites need
not be equally spaced. We prove that all four chains are solvable
by deriving an exact expression for their partition function using
Polychronakos's ``freezing trick''. {}From this expression we
deduce several properties of the spectrum, and advance a number of
conjectures that hold for a wide range of values of the spin $M$
and the number of particles. In particular, we conjecture that the
level density is Gaussian, and provide a heuristic derivation of
general formulas for the mean and the standard deviation of the
energy.
\end{abstract}

\begin{keyword}
Spin chains \sep exact solvability \sep integrability \sep
Calogero--Sutherland models \sep Dunkl operators

  \PACS 75.10.Pq \sep 03.65.Fd
\end{keyword}
\end{frontmatter}

\section{Introduction}\label{intro}

The Haldane--Shastry (HS) chain~\cite{Ha88,Sh88} describes a fixed
arrangement of equally spaced spin $1/2$ particles in a circle
with pairwise interactions inversely proportional to the square of
the chord distance between the particles. The original interest of
this model lies in the fact that the $U\to\infty$ limit of
Gutzwiller's variational wave function for the Hubbard
model~\cite{Gu63,GV87,GJR87}, which also coincides with the
one-dimensional version of the resonating valence bond state
introduced by Anderson~\cite{ABZH87}, is an exact eigenfunction of
the HS chain. The exact solvability of the HS chain was already
proved in the original papers of Haldane and Shastry. A few years
later Fowler and Minahan~\cite{FM93} used Polychronakos's
exchange-operator formalism~\cite{Po92} to show that this model
is also completely integrable. Although the obvious relation of
the HS chain with the Sutherland (scalar) model of $A_N$
type~\cite{Su71,Su72,OP83} was already remarked by Shastry, an
explicit quantitative connection was first established by
Polychronakos through the so-called ``freezing
trick''~\cite{Po93}. In the latter paper it is shown how to
construct an integrable spin chain from a
Calogero--Sutherland (CS) model of $A_N$ type with internal degrees of freedom
(``spin'')~\cite{HH92,BGHP93,HW93,MP93,SS93} by freezing the
particles at the classical equilibrium positions of the scalar
part of the CS potential. The first integrals of the spin chain
are essentially obtained as the large coupling constant limit of
the first integrals of the corresponding CS model. Polychronakos
applied this technique to the original (rational) Calogero model
of $A_N$ type~\cite{Ca71}, constructing in this way a new
integrable spin chain of HS type in which the spin sites were no
longer equally spaced. In a subsequent publication~\cite{Po94},
the same author gave a heuristic argument based on the freezing
trick that relates the spectrum of the integrable spin chain with
those of the corresponding scalar and spin dynamical models.

Both the integrability and the spectrum of the Haldane--Shastry
and Polychronakos spin chains can thus be obtained from the
trigonometric and rational CS spin dynamical models of $A_N$ type.
By contrast, the spin chains associated with the spin models of
$BC_N$ type~\cite{Ya95,YT96,Du98,FGGRZ01b,IS01,FGGRZ03,CC04} have
received comparatively little attention. This is in part due to
the fact that, unlike their $A_N$ counterparts, the $BC_N$-type
spin chains depend nontrivially on free parameters (one in the
rational case and two in the trigonometric or hyperbolic cases).
The integrability of the spin chain associated with the $BC_N$
rational CS model was established by Yamamoto and Tsuchiya
\cite{YT96} using the Dunkl operator formalism
\cite{Po92,Du89,Ch91}, although, to the best of the authors'
knowledge, the spectrum of this model has not been computed so
far. The Haldane--Shastry (trigonometric) spin chain of $BC_N$
type was discussed by Bernard, Pasquier, and Serban \cite{BPS95},
but only for spin $1/2$ and with the assumption that the sites are
equally spaced, which restricts the pair of free parameters in the
model to just three particular values. Finkel et
al.~\cite{FGGRZ03} recently discussed the integrability of the
hyperbolic HS spin chain of $BC_N$ type, but did not examine its
spectrum.

In this paper we study the $BC_N$ version of the Haldane--Shastry
spin chain for arbitrary values of the spin and the coupling
constants. It turns out that there are actually four different
$BC_N$ spin chains related to the original Haldane--Shastry chain,
two of which are ferromagnetic and the other two
antiferromagnetic. We prove that these chains are exactly solvable
provided that the sites are the coordinates of an equilibrium of a
suitable scalar potential, which is the same for all four chains.
In particular, for generic values of the coupling constants the
sites are not equally spaced. In addition, we rigorously establish
the essential uniqueness of the equilibrium point of the scalar
potential determining the chain sites. Using Polychronakos's
freezing trick, we are able to derive an exact expression for the
partition function of the models, thus establishing their
solvability. {}From this expression, which is the main result of
this paper, we deduce several interesting general properties of
the spectrum. In the first place, the spectrum depends on the
coupling constants only through their semisum $\Bbe$, while for
generic values of $\Bbe$ the degeneracies of the energy levels
depend only on the spin $M$ and the number of particles $N$.
Secondly, although the energy levels are in general unequally
spaced, for $\Bbe\gg N$ (and sufficiently large $M$) they cluster
around an equally spaced set. In the third place, for half-integer
spin the spectra of the two types of (anti)ferromagnetic chains
are exactly the same, even if their Hamiltonians differ by a
nontrivial term.

Apart from the rigorous results just mentioned, the evaluation of
the partition function for several values of $M$ and $N$ has led
us to several conjectures regarding the spectrum. First of all,
our calculations strongly suggest that the clustering of the
levels around an equally spaced set when $\Bbe\gg N$ occurs in
fact for all values of the spin $M$. Secondly, even for moderately
large values of $N$ the level density follows a Gaussian
distribution with great accuracy. This fact, which is the main
conjecture of this paper, is reminiscent of the analogous property
of the ``embedded Gaussian ensemble'' (EGOE) in Random Matrix
Theory~\cite{MF75}. It should be noted, however, that the
essential requirement defining the EGOE, namely that the ratio of
the number of particles to the number of one-particle states tend
to zero as both quantities tend to infinity does not hold in our
case. If the level density is Gaussian to a very high degree of
approximation (for sufficiently large $N$), it is fully
characterized by the mean $\mu$ and standard deviation $\si$ of
the energy. {}From a natural conjecture on the dependence of $\mu$
and $\sigma$ on the number of particles, we have derived general
formulas expressing these parameters as functions of $N$ and $M$.
We have then checked that these formulas yield the exact values of
$\mu$ and $\si$ for a wide range of values of the spin and the
number of particles. We have also rigorously proved (without
making use of the previous conjectures) that the standard
deviations for both types of (anti)ferromagnetic chains with
integer spin exactly coincide, even if their spectra are
essentially different.

The paper is organized as follows. In Section~\ref{intSuth} we
introduce the Sutherland model of $BC_N$ type with internal
degrees of freedom, and outline a proof of its integrability by
expressing the Hamiltonian in terms of an appropriate commuting
family of self-adjoint Dunkl operators. The spectrum of the latter
model is determined in Section~\ref{specSuth} by explicit
triangularization of the Hamiltonian. In particular, we compute
the ground state energy in terms of the parameters of the model.
The four types of Haldane--Shastry spin chains of $BC_N$ type,
which are the main subject of this paper, are presented in
Section~\ref{intHS}. Section~\ref{speHS} is devoted to the
calculation of the spectrum of the chains introduced in the
previous section. We first provide a semi-rigorous detailed
justification of the freezing trick, whose key points are the
uniqueness of the equilibrium point of the associated scalar
potential together with the knowledge of the full spectrum of the
corresponding scalar and spin Sutherland models of $BC_N$ type.
{}From the freezing trick we directly obtain an explicit
expression for the ground state energy of the chains. We next make
use of the freezing trick (which, by itself, does not completely
determine the spectrum) to compute in closed form the partition
functions of all four types of $BC_N$ spin chains. In the last
section we present concrete examples for spin $1/2$ and $1$, which
led us to formulate the general conjectures mentioned above. The
paper ends with a technical appendix, in which we establish the
uniqueness of the equilibrium point of the scalar potential
determining the sites of the chains.

\section{The spin dynamical models}\label{intSuth}

In this section we shall study the integrability of the
\emph{trigonometric Sutherland spin models of $BC_N$ type}. Each of
these models describes a system of $N$ identical particles with
internal degrees of freedom (``spin'') moving on a circle, subject to one- and
two-body interactions depending on the particles' spatial and internal
coordinates. We shall denote by $\cS$ the finite-dimensional Hilbert
space corresponding to the spin degrees of freedom spanned by the states
$|s_1,\dots,s_N\rangle$, where $-M\le s_i\le M$ and $M$ is a half-integer.
We shall respectively denote by $S_{ij}$ and $S_i$ ($i,j=1,\ldots,N$)
the spin permutation and reversal operators, whose action on the basis
of spin states is defined by
\begin{equation}
\begin{aligned}
  & S_{ij}|s_1,\dots,s_i,\dots,s_j,\dots,s_N\rangle=|s_1,\dots,
  s_j,\dots,s_i,\dots,s_N\rangle\,,\\
  &
  S_i|s_1,\dots,s_i,\dots,s_N\rangle=|s_1,\dots,-s_i,\dots,s_N\rangle\,.
\label{SS}
\end{aligned}
\end{equation}
These operators are represented in $\cS$ by $(2M+1)^N$-dimensional
Hermitian matrices. We shall denote by $\fS$ the multiplicative group
generated by the operators $S_{ij}$ and $S_i$, which is isomorphic to
the Weyl group of $B_N$ type. We shall also use the customary notation
$\tS_{ij}=S_iS_jS_{ij}$.

The $BC_N$-type spin dynamical models we shall study in this
section are collectively described by a Hamiltonian of the form
\begin{equation}
\label{Hstar}
\begin{aligned}
  H^*_{\ep\ep'}=-\sum_i \pa_{x_i}^2 &+ a\,\sum_{i\neq j}\left[\sin^{-2}
    x_{ij}^-\,(a-\ep\,S_{ij})+\sin^{-2} x_{ij}^+\,(a-\ep\,\tS_{ij})\right]\\
  &{}+b\,\sum_i \sin^{-2}\!x_i\,(b-\ep' S_i)+b'\,\sum_i
  \cos^{-2}\!x_i\,\big(b'-\ep' S_i\big)\,,
\end{aligned}
\end{equation}
where $\ep,\ep'=\pm1$ are two independent signs, $a$, $b$, $b'$
are real parameters greater than $1/2$, and $x_{ij}^\pm=x_i\pm
x_j$. Here and in what follows, the sums (and products) run from
$1$ to $N$ unless otherwise constrained. The potential
in~\eqref{Hstar} possesses inverse-square type singularities at
the hyperplanes $x_i\pm x_j=k\pi$, $x_i=k\pi/2$, with $k\in\ZZ$.
In fact, since the nature of these singularities makes it
impossible for one particle to overtake another or to cross the
singularities at $x_i=k\pi/2$, we can regard the particles as
distinguishable and take as configuration space the set
\begin{equation}\label{tC}
\tC=\Big\{\bx=(x_1,\ldots,x_N)\in \RR^N\;\Big|\;0<x_1<\cdots
<x_N<\frac\pi2\Big\}\,.
\end{equation}
The Hilbert space of the system may thus be taken as
$\cH=L^2_0(\tC)\otimes\cS$, where
\begin{multline*}
  L^2_0(\tC)=\Big\{f\in L^2(\tC)\;\Big|\;\exists\lim_{x_i\pm x_j\to k\pi}
  |x_i\pm x_j-k\pi|^{-a}|f(\bx)|\,,\quad\exists\lim_{x_i\to 0} |x_i|^{-b}|f(\bx)|\,,\\
  \exists\lim_{x_i\to\pi/2} |x_i-\pi/2|^{-b'}|f(\bx)|\,; \quad
  k=0,1,\quad 1\leq i\neq j\leq N\Big\}.
\end{multline*}
Note, in particular, that the physical
wavefunctions vanish faster than the square root of the distance to
the singular hyperplanes in their vicinity.

Formally, the four Hamiltonians~\eqref{Hstar} can be represented as a single
Hamiltonian $H^*=H^*_{\ep\ep'}$ for an arbitrary choice of the signs $\ep$ and
$\ep'$, provided that the parameters $a$, $b$ and $b'$ are also allowed
to take negative values less than $-1/2$. We have preferred to use the more
explicit representation~\eqref{Hstar} since, as we shall see in
the following section, the spectrum of $H^*_{\ep\ep'}$
depends in an essential way on $\ep$ and $\ep'$.
It can be shown that the operator
$H^*_{\ep\ep'}:\cH\ra\cH$ is equivalent to any of its extensions to
spaces of symmetric or antisymmetric functions (with respect to both
permutations and sign reversals) in $L_0^2(C)\otimes\cS$, where $C$ is
the $N$-cube $(-\frac\pi2,\frac\pi2)^N$ and $L_0^2(C)$ is defined
similarly to $L^2_0(\tC)$. We shall consider without loss of generality
that $H^*_{\ep\ep'}$ acts in the Hilbert space
\begin{equation}
\cH_{\ep\ep'}=\La_{\ep\ep'}\big(L^2_0(C)\otimes\cS\big)\,,
\end{equation}
where $\La_{\ep\ep'}$ is the projection operator on states with
parity $\ep$ under simultaneous permutations of spatial
coordinates and spins and $\ep'$ under sign reversals. The latter
operator is characterized by the relations
\begin{equation}\label{La}
K_{ij}\La_{\ep\ep'}=\ep\,S_{ij}\La_{\ep\ep'}\,,\qquad
K_{i}\La_{\ep\ep'}=\ep'S_{i}\La_{\ep\ep'}\,,
\end{equation}
where $K_{ij}$ and $K_i$ respectively denote the spatial
coordinates' permutation and sign reversing operators, defined by
\begin{align*}
&(K_{ij}f)(x_1,\dots,x_i,\dots,x_j,\dots,x_N)=f(x_1,\dots,
x_j,\dots,x_i,\dots,x_N)\,,\\
&(K_i f)(x_1,\dots,x_i,\dots,x_N)=f(x_1,\dots,-x_i,\dots,x_N)\,.
\end{align*}

The relations~\eqref{La} suggest the definition of a mapping
${}^*_{\ep\ep'}:\fD\otimes\fK\to\fD\otimes\fS$, where $\fD$ denotes the algebra
of scalar linear differential operators and
$\fK\simeq\fS$ is the multiplicative group generated by the operators
$K_{ij}$ and $K_i$, as follows:
\begin{equation}\label{star}
\big(D K_{i_1j_1}\cdots K_{i_rj_r}K_{l_1}\cdots K_{l_s}\big)^*_{\ep\ep'}
=\ep^r{\ep'}^s
D\,S_{l_s}\cdots S_{l_1}S_{i_rj_r}\cdots S_{i_1j_1}\,,
\end{equation}
where $D\in\fD$.  This determines a linear map $A\mapsto A^*$ in
$\fD\otimes\fK$, which by Eq.~\eqref{La} satisfies
\begin{equation}\label{AA*}
A\,\La_{\ep\ep'}=A^*_{\ep\ep'}\La_{\ep\ep'}.
\end{equation}
In particular, each of the
physical Hamiltonians $H^*_{\ep\ep'}$ in~\eqref{Hstar} is the image
under the corresponding star mapping of a \emph{single} operator $H$,
given by
\begin{equation}
\label{H}
\begin{aligned}
  H=-\sum_i \pa_{x_i}^2 &+ a\,\sum_{i\neq j}\left[\sin^{-2}
    x_{ij}^-\,(a-K_{ij})+\sin^{-2} x_{ij}^+\,(a-\tK_{ij})\right]\\
  &{}+b\,\sum_i \sin^{-2}\!x_i\,(b-K_i)+b'\,\sum_i
  \cos^{-2}\!x_i\,\big(b'-K_i\big)\,.
\end{aligned}
\end{equation}

The integrability of the Hamiltonian~\eqref{Hstar} can be
established by the same method applied in~Ref.~\cite{FGGRZ03} to
the hyperbolic version of $H^*_{--}$, based on the
fact that $H$ can be expressed as the sum
of the squares of the commuting Dunkl operators
\begin{multline}\label{J}
  J_k=\iu\,\pa_{x_k}+a\sum_{l\neq k}\Big[(1-\iu\cot
  x_{kl}^-)\,K_{kl}+(1-\iu\cot x_{kl}^+)\,\tK_{kl}\Big]\\
  +\big[b\,(1-\iu\cot x_k)+b'\,(1+\iu\tan
  x_k)\big]K_k-2a\sum_{l<k}K_{kl}\,.
\end{multline}
These operators are related to the hyperbolic Dunkl operators $\hJ_k$
of Ref.~\cite{FGGRZ03} by $J_k(\bx)=-\hJ_k(\iu\bx)$.
The commutativity of the Dunkl operators $J_k$ implies that the operators
\begin{equation}\label{Ip}
I_p=\sum_k J_k^{\,2p}\,,\qquad p=1,\ldots,N\,,
\end{equation}
form a complete set of commuting integrals of motion of $H=I_1$.
{}From this fact it follows, as in Ref.~\cite{FGGRZ03}, that the
corresponding operators $(I_p)^*_{\ep\ep'}$, $p=1,\ldots,N$, act
on the Hilbert space $\cH_{\ep\ep'}$ and form a complete set of
integrals of motion of the Hamiltonian $H^*_{\ep\ep'}=(I_1)^*_{\ep\ep'}$.

To end this section, we shall prove that the integrals of motion
$(I_p)^*_{\ep\ep'}$ are self-adjoint. Note first of all that,
unlike the operators $\hJ_k$, the Dunkl operators~\eqref{J} and
hence the integrals of motion~\eqref{Ip} are self-adjoint. Since,
furthermore, $I_p$ and $\La_{\ep\ep'}$ are self-adjoint and
commute with one another (in fact, $I_p$ commutes with $K_{ij}$
and $K_i$ by Lemma~4 of Ref.~\cite{FGGRZ03}), we have
$$
(I_p)^*_{\ep\ep'}\La_{\ep\ep'}
=I_p\La_{\ep\ep'}
=(I_p\La_{\ep\ep'})^\dagger
=\big((I_p)^*_{\ep\ep'}\La_{\ep\ep'}\big)^\dagger\,.
$$
On the other hand, since $(I_p)^*_{\ep\ep'}$ also commutes with
$\La_{\ep\ep'}$ we obtain
$$
\big((I_p)^*_{\ep\ep'}\La_{\ep\ep'}\big)^\dagger
=\big(\La_{\ep\ep'} (I_p)^*_{\ep\ep'}\big)^\dagger
={\vphantom{\big|}(I_p)^{*}_{\ep\ep'}}^{\!\!\!\dagger}\,\La_{\ep\ep'}\,,
$$
from which it follows that
${\vphantom{\big|}(I_p)^{*}_{\ep\ep'}}^{\!\!\!\dagger}
=(I_p)^*_{\ep\ep'}$ by Lemma 1 of Ref.~\cite{FGGRZ03}.

\section{Spectrum of the spin dynamical models}\label{specSuth}

In this section we shall compute the spectrum of the trigonometric
Sutherland spin models of $BC_N$ type \eqref{Hstar}. The results
of this section will be used in Section~\ref{speHS}
to derive the asymptotic behavior of the partition
function of these models in the large coupling
constant limit.

The computation of the spectrum of the Hamiltonian~\eqref{Hstar} is
analogous to the corresponding computation for the hyperbolic model
studied in Ref.~\cite{FGGRZ03}, in spite of the fact that the boundary
conditions are different. The starting point of this computation is
the invariance under the Dunkl operators $J_i$ of the
finite-dimensional spaces
\begin{equation}\label{Rdef}
\cR_k = \big\langle\phi(\bx)\exp\big(2\iu\sum_j n_j
x_j\big)\;\big|\; n_j=-k,-k+1,\dots, k\,,\quad j=1,\dots,N
\big\rangle\,,
\end{equation}
where
\begin{equation}\label{phi}
\phi(\bx) = \prod_{i<j}|\sin x_{ij}^-\, \sin x_{ij}^+|^a\cdot
\prod_i|\sin x_i|^b |\cos x_i|^{b'}\,,
\end{equation}
for all nonnegative integer values of $k$. It follows
that the operator $H=I_1$ preserves the spaces $\cR_k$ for all $k$. Since
$H$ commutes with $\La_{\ep\ep'}$, Eq.~\eqref{AA*} implies that
\begin{equation}\label{HstH}
H^*_{\ep\ep'}\big[\La_{\ep\ep'}\big(\vp\ket\si\big)\big] =
\La_{\ep\ep'}\big[(H\vp)\ket\si\big]\,,
\end{equation}
for all $\vp\in L^2_0(C)$ and $\ket\si\in\cS$. Hence the Hamiltonian
$H^*_{\ep\ep'}$ leaves invariant the infinite increasing sequence of
finite-dimensional spaces
\begin{equation}
\cM_{k,\ep\ep'}= \La_{\ep\ep'}(\cR_k\otimes\cS)\,,\qquad k=0,1,\dots,
\end{equation}
and is therefore exactly solvable in the sense of Turbiner~\cite{Tu88,Tu92}.

We shall next construct a (non-orthonormal) basis $\cB$ of the Hilbert
space $L^2_0(C)$ in which $H$ is represented by a triangular
infinite-dimensional matrix, thereby obtaining an exact formula for
the spectrum of this operator. To this end, note that the (scaled)
exponential monomials
\begin{equation}\label{fn}
f_n(\bx) = \phi(\bx)\exp\big(2\iu\sum_j n_j x_j\big)\,,\qquad
n=(n_1,\dots,n_N)\,,\quad n_j\in\ZZ\,,
\end{equation}
span a dense subspace of the Hilbert space $L^2_0(C)$. We can
introduce a partial ordering $\prec$ in the set of exponential
monomials~\eqref{fn} as follows. Given a multiindex
$n=(n_1,\dots,n_N)\in\ZZ^N$, we define the nonnegative and
nonincreasing multiindex $[n]$ by
\begin{equation}\label{brn}
[n]=(\abs{n_{i_1}},\dots,\abs{n_{i_N}})\,, \qquad\text{where}\quad
\abs{n_{i_1}}\geq\dots\geq\abs{n_{i_N}}\,.
\end{equation}
If $n,n'\in[\ZZ^N]$ are nonnegative and nonincreasing multiindices, we
shall say that $n\prec n'$ if $n_1-n_1'=\dots=n_{i-1}-n_{i-1}'=0$ and
$n_i <n_i'$. For two arbitrary multiindices $n,n'\in\ZZ^N$, by
definition $n\prec n'$ if and only if $[n]\prec[n']$. Finally, we
shall say that $f_n\prec f_{n'}$ if and only if $n\prec n'$. Note that
the partial ordering $\prec$ is preserved by the action of the Weyl
group $\fK$, i.e., if $f_n\prec f_{n'}$ then $Wf_n\prec Wf_{n'}$ for
all $W\in\fK$.

We can take as the basis $\cB$ any ordering of the set of exponential
monomials~\eqref{fn} compatible with the partial ordering $\prec$.
This follows from the fact that
\begin{equation}\label{Hgfn}
H f_n = \sum_i\la_{[n],i}^2 f_n + \sum_{\substack{n'\in \ZZ^N\\
n'\prec \,n}}c_n^{n'}\,f_{n'}\,,\qquad n\in\ZZ^N,
\end{equation}
where $\la_{[n],i}$ and $c_n^{n'}$ are real numbers
(cf.~Proposition~2 of Ref.~\cite{FGGRZ03}).  The numbers $\la_{m,i}$
($m\in[\ZZ^N]$) are explicitly given by
\begin{equation}\label{la}
\la_{m,i} =
\begin{cases}
  2m_i+b+b'+2a\big(N+i+1-\#(m_i)-2\ell(m_i)\big),\quad& m_i>0\,,\\[3pt]
  -b-b'+2a(i-N)\,,\quad& m_i=0\,,
\end{cases}
\end{equation}
where we have used the following notation:
\[
\#(s)=\card\{i\:|\:m_i=s\}\,,\qquad \ell(s)=\min\{i\:|\:m_i=s\}\,.
\]
For instance, if $m=(5,2,2,1,1,1,0)$ then $\#(1)=3$ and
$\ell(1)=4$. It will also be convenient in what follows to take
$\ell(s)=+\infty$ if $m_i\neq s$ for all $i=1,\dots,N$.
Equation~\eqref{Hgfn} implies that the operator $H$ is represented
in the basis $\cB$ by an upper triangular matrix with diagonal
elements
\[
E_n=\sum_i\la_{[n],i}^2\,.
\]
{}From the previous formula it is straightforward to deduce the
following more compact expression for the eigenvalues $E_n$ of the
operator~$H$:
\begin{equation}\label{En}
E_n=\sum_i\big(2[n]_i+b+b'+2a(N-i)\big)^2\,.
\end{equation}
Indeed, if $m=[n]\in [\ZZ^N]$ and
$m_{k-1}>m_k=\cdots=m_{k+p}>m_{k+p+1}\geq 0$ then $\ell(m_{k+j})=k$ and
$\#(m_{k+j})=p+1$ for $j=0,\ldots,p$, so that
\[
\la_{m,k+j}=2 m_{k+j}+b+b'+2a(N-k-p+j)=
2m_{k+p-j}+b+b'+2a\big(N-(k+p-j)\big)
\]
and hence
\begin{equation}\label{partialsum}
\sum_{i=k}^{k+p}\la_{m,i}^2=\sum_{i=k}^{k+p}\big(2m_i+b+b'+2a(N-i)\big)^2\,.
\end{equation}
If, on the other hand, $m_{k-1}>m_k=\cdots=m_N=0$,
Eq.~\eqref{partialsum} follows directly from~\eqref{la}. This
completes the proof of the formula~\eqref{En}.

Let us now compute the spectrum of the Hamiltonian~\eqref{Hstar} in
$\cH_{\ep\ep'}$. Note, first of all, that the states of the form
\begin{equation}\label{monbasis}
\La_{\ep\ep'}\big(f_n\ket{s_1,\dots,s_N}\big)\,,\qquad n\in [\ZZ^N]\,,
\end{equation}
span a dense subset of the Hilbert space $\cH_{\ep\ep'}$, by the
analogous property of the functions~\eqref{fn} in $L_0^2(C)$. The
states~\eqref{monbasis}, however, are not linearly independent (in
particular, some of them vanish if $\ep$ or $\ep'$ are negative).
A (non-orthonormal) basis of
$\cH_{\ep\ep'}$ may be obtained from the states~\eqref{monbasis} by
imposing the following conditions on the spin vector
$\ket{s_1,\dots,s_N}$ (cf.~Proposition~3 of Ref.~\cite{FGGRZ03}):
\begin{subequations}\label{conditions}
\begin{alignat}{2}
  \text i&)\quad s_i-s_j\geq\de_{-1,\ep}\,,\quad &&\text{if}\quad
  n_i=n_j\quad\text{and}\quad i<j\,;
  \label{cond2}\\[1mm]
  \text{ii}&)\quad s_i\geq\frac12\,\de_{-1,\ep'}\,,&&\text{if}\quad
  n_i=0\,,\label{cond3}
\end{alignat}
\end{subequations}
where $\de$ is Kronecker's delta. Indeed, if $n_i=n_j$ with $i<j$ we
can clearly permute the $i$-th and $j$-th particles (if necessary) so
that $s_i\geq s_j$, leaving the state~\eqref{monbasis} invariant up to
a sign. If, in addition, $\ep=-1$ we must have $s_i>s_j$ by
antisymmetry under permutations. Likewise, if $n_i=0$ we can assume
that $s_i\geq 0$ after a possible reversal of the sign of the
coordinates of the $i$-th particle, which again preserves the
state~\eqref{monbasis} up to a sign. Moreover, $\ep'=-1$ forces
$s_i>0$ by antisymmetry under sign reversals. Note that when $\ep=-1$,
i.e., when the basis states~\eqref{monbasis} are antisymmetric with
respect to permutations, the first condition implies the following
restriction on the multiindex $n\in[\ZZ^N]$:
\begin{equation}
  \#(n_i)\leq\begin{cases}
    2M+1\,,& \text{if}\qquad n_i>0\\[1mm]
    \hfill M_{\ep'}\,,\hfill&
    \text{if}\qquad n_i=0\,,
\end{cases}\label{cond1}
\end{equation}
where $M_+=\lfloor M\rfloor+1$ and $M_-=\lceil M\rceil$. Here $\lfloor x\rfloor$
and $\lceil x\rceil$ denote respectively the integer part of $x$ and the smallest
integer greater than or equal to $x$.
Let $\cB_{\ep\ep'}$ be any ordering of the set of
states~\eqref{monbasis}--\eqref{cond1} compatible with the partial
ordering $\prec$. It follows from Eqs.~\eqref{HstH}
and~\eqref{Hgfn} that the matrix of $H^*_{\ep\ep'}$ with respect
to the basis $\cB_{\ep\ep'}$ is upper triangular, with eigenvalues
given by
\begin{equation}\label{Enstar}
E^*_{\ep\ep'}(n;s)=\sum_i\big(2n_i+b+b'+2a(N-i)\big)^2\,,
\qquad n\in[\ZZ^N]\,,
\quad s=(s_1,\dots,s_N)\,.
\end{equation}
It is worth mentioning at this point that, although a cursory
inspection of the previous equation may suggest that the
models~\eqref{Hstar} are isospectral, this is in general not the
case. Indeed, condition~\eqref{cond1} implies that many
eigenvalues of the models with $\ep=1$ are absent from the
spectrum of the models with $\ep=-1$. Besides, for a fixed value
of $\ep$, by condition~\eqref{cond3} the degeneracy of the
eigenvalues also depends on $\ep'$ when $M$ is an integer (see
Eqs.~\eqref{nm} and \eqref{dk} in Section~\ref{speHS} for the
minimum degeneracy of each level).

Since $E_{\ep\ep'}^*(n;s)$ is an increasing function of the components of the
multiindex $n$, the ground state of the system is obtained when each component
$n_i$ takes the lowest possible value. Thus for $\ep=1$, or $\ep=-1$ and $N\leq
M_{\ep'}$, we have $n=0$. On the other hand, when $\ep=-1$ and $N>
M_{\ep'}$ condition~\eqref{cond1} implies that
\begin{equation}
n = \big(\overbrace{\vphantom{1}m_0,\dots,m_0}^r,
\overbrace{m_0-1,\dots,m_0-1}^{2M+1}
,\dots,\overbrace{1,\dots,1}^{2M+1},
\overbrace{0,\dots,0}^{M_{\ep'}}\big),
\label{n}
\end{equation}
where $N=M_{\ep'}+(m_0-1)(2M+1)+r$ with $r=1,\ldots,2M+1$. The
ground state energy $E^*_{\ep\ep'\!,\text{min}}$ is easily
computed in this case using Eq.~\eqref{Enstar} with the multiindex
$n$ given in~\eqref{n}. We thus obtain
\begin{align}
E^*_{\ep\ep'\!,\text{min}}&=\frac43\,a^2N^3
-2\ms acN^2+\frac13\,(3c^2-a^2)N+
\frac13\,\kappa\,m_0\big[4m_0^2(1-a\ms\kappa)\notag\\
&\qquad+6\ms c\ms m_0+a\ms\kappa+2\big] +2\ms
m_0\rho\big[c+m_0(1-a\ms\kappa)-\frac12\,a\ms\rho\big]\,,\label{Emin}
\end{align}
where $c=a-b-b'-2m_0$ and
\begin{equation}\label{murho}
\kappa=2M+1\,,\qquad\rho=
\begin{cases}
\ep', & M=0,1,\ldots\\[1mm]
0, & M=\dfrac12\,,\dfrac32\,,\,\ldots\,.
\end{cases}
\end{equation}
It can be easily shown that Eq.~\eqref{Emin} is also
valid when $n=0$ if we take $m_0=0$. Thus Eq.~\eqref{Emin} yields
the ground state energy in all cases provided that $m_0$ is
defined by
\begin{equation}\label{mzero}
m_0=\de_{\ep,-1}\bigg\lceil\frac{N-M_{\ep'}}{2M+1}\bigg\rceil\,.
\end{equation}

\section{The spin chains}\label{intHS}

The Hamiltonian of the HS spin chains of $BC_N$ type associated
with the spin dynamical models \eqref{Hstar} discussed in the
previous sections is given by
\begin{multline}\label{ssh}
\ssh_{\ep\ep'}=\sum_{i\neq j}\Big[\sin^{-2}\xi_{ij}^-\,(1-\ep S_{ij})
+\sin^{-2}\xi_{ij}^+\,(1-\ep\tS_{ij})\Big]\\
{}+\sum_i\big(\be\,\sin^{-2}\xi_i+\be'\,\cos^{-2}\xi_i\big)(1-\ep'S_i)\,,
\end{multline}
where $\be$ and $\be'$ are positive real parameters,
$\xi_{ij}^\pm=\xi_i\pm\xi_j$, and $\bxi=(\xi_1,\ldots,\xi_N)$ is the unique
equilibrium point in the set $\tC$ of the classical potential
\begin{equation}\label{U}
U(\bx)=\sum_{i\neq
  j}(\sin^{-2}x_{ij}^-+\sin^{-2}x_{ij}^+)+\sum_i(\be^2\sin^{-2}x_i+\be'^2\cos^{-2}x_i)\,.
\end{equation}
It is important to note that the classical potential \eqref{U} is
independent of $\ep$ and $\ep'$, and therefore the sites of the four chains
\eqref{ssh} are the same.  The existence of a minimum of $U$ in $\tC$ for all
values of $\be$ and $\be'$ is a consequence of the positivity and continuity
of $U$ in $\tC$ and the fact that it tends to infinity at the boundary of this
set.  The uniqueness of this minimum is proved in Appendix~\ref{AppB}.  Note
that, in contrast, the corresponding potential for the hyperbolic spin chain
of $BC_N$ type treated in~\cite{FGGRZ03} admits an equilibrium point only for
a certain range of values of $\be$ and $\be'$.

The chains~\eqref{ssh} with $\ep=-1$ (respectively $\ep=1$) are of
antiferromagnetic (respectively ferromagnetic) type. Note also that
the spin chain Hamiltonians $\ssh_{-\ep,-\ep'}$ and
$-\ssh_{\ep\ep'}$ are related by
\begin{equation}\label{hahf}
\ssh_{-\ep,-\ep'}=-\ssh_{\ep\ep'}+2V(\bxi),
\end{equation}
where
\begin{equation}\label{V}
V(\bx)= \sum_{i\neq j}\left(\sin^{-2} x_{ij}^-+ \sin^{-2} x_{ij}^+\right) +\be\,\sum_i
\sin^{-2}\!x_i+\be'\,\sum_i \cos^{-2}\!x_i\,.
\end{equation}
On the other hand, for a fixed $\ep$ the two chains
$\ssh_{\ep,\pm}$ are essentially different. {}From Eq.~\eqref{ssh}
it immediately follows that the eigenvalues $\sse_{\ep\ep',j}$ of
the Hamiltonian $\ssh_{\ep\ep'}$ are nonnegative. Moreover, in the
ferromagnetic case ($\ep=1$) the ground state energy clearly
vanishes, since states symmetric under permutations with parity
$\ep'$ under spin reversals are annihilated by the Hamiltonian.
Equation~\eqref{hahf} implies that the maximum energy of the
antiferromagnetic chains is $2V(\bxi)$.

The integrability of the spin chain~\eqref{ssh} with $M=1/2$ and
$\ep=\ep'=-1$ was proved in Ref.~\cite{BPS95} only for the
special values $(3/2,1/2)$, $(3/2,3/2)$, and $(1/2,1/2)$ of the pair $(\be,\be')$, for which
the corresponding sites
\[
\frac{i\pi}{2N+1}\,,\quad \frac{i\pi}{2N+2}\,,\quad \Big(i-\frac12\Big)\,\frac{\pi}{2N}\,;\qquad
i=1,\dots,N,
\]
are \emph{equally spaced}, as in the original Haldane--Shastry chain~\cite{Ha88,Sh88}.
As we shall see below, the discussion of the integrability of the
HS spin chains of $BC_N$ type~\eqref{ssh} for arbitrary values of the
parameters $\be$ and $\be'$ is completely analogous to that of
Ref.~\cite{FGGRZ03} for the hyperbolic version of $\ssh_{--}$.

Let us begin by defining the operators
$\ssJ_i$ ($i=1,\dots, N$) and $\ssI_p$ ($p\in\NN$) by
$$
J_i = \iu\,\pa_{x_i}+a\ssJ_i\,,\qquad \ssI_p=\sum_i\ssJ_i^{2p}.
$$
The operators $(\ssI_p)^*_{\ep\ep'}$ clearly commute with one another, since
$[(\ssI_p)^*_{\ep\ep'},(\ssI_q)^*_{\ep\ep'}]$ is the coefficient of
$a^{2(p+q)}$ in the expansion in powers of $a$ of the identity
$[(I_p)^*_{\ep\ep'},(I_q)^*_{\ep\ep'}]=0$.  Hence the operators
\begin{equation}
(\ssI_p)^{*0}_{\ep\ep'}\equiv
\left.(\ssI_p)^*_{\ep\ep'}\right|_{\bx=\bxi}\,,\qquad
p\in\NN\,,
\label{Ipast0}
\end{equation}
also commute with one another. We shall now prove that the
operators \eqref{Ipast0} form a commuting family of integrals of
motion for the spin chain Hamiltonian $\ssh_{\ep\ep'}$. Note that
this result does not follow trivially from the previous assertions
since, in contrast with the dynamical case,
$(\ssI_1)^{*0}_{\ep\ep'}=U(\bxi)$ is a constant and therefore does
not coincide with $\ssh_{\ep\ep'}$.

The starting point in the proof of the commutativity of $\ssh_{\ep\ep'}$ and
$(\ssI_p)^{*0}_{\ep\ep'}$ is the following expansion of $H$ in powers of $a$:
\begin{equation}
\label{sH2}
H=-\sum_i\pa_{x_i}^2-a\,\ssH + a^2 U(\bx)\,,
\end{equation}
where
\begin{equation}
\label{ssH}
\ssH=\sum_{i\neq j}\Big[\sin^{-2}x_{ij}^-\,K_{ij}
+\sin^{-2}x_{ij}^+\,\tK_{ij}\Big]
{}+\sum_i\big(\be\,\sin^{-2}x_i+\be'\,\cos^{-2}x_i\big)K_i\,,
\end{equation}
$U$ is defined in Eq.~\eqref{U}, and we have set
\begin{equation}\label{bebep}
\be=\frac ba\,,\qquad\be'=\frac{b'}a\,.
\end{equation}
Note that $\ssh_{\ep\ep'}=-\ssH^{*0}_{\ep\ep'}+V(\bxi)$, so that the
integrability of $\ssh_{\ep\ep'}$ follows from that of $\ssH^{*0}_{\ep\ep'}$.
Arguing as in Ref.~\cite{FGGRZ03} it is straightforward to show that
$$
[\ssH^*_{\ep\ep'},(\ssI_p)^*_{\ep\ep'}]=
\sum_i \frac{\pa U}{\pa x_i}\,(\ssC_{p,i})^*_{\ep\ep'}\,,
$$
for certain operators $\ssC_{p,i}$ in $\fD\otimes\fK$. Setting
$\bx=\bxi$ in the previous identity, it follows that
$(\ssI_p)^{*0}_{\ep\ep'}$ commutes with $\ssh_{\ep\ep'}$. Note
finally that the first integrals $(\ssI_p)^{*0}_{\ep\ep'}$ are
clearly self-adjoint, since they are equal to the coefficient of
$a^{2p}$ in the corresponding self-adjoint operators
$(I_p)^*_{\ep\ep'}$ evaluated at the equilibrium point $\bxi$.

\section{Partition function and spectrum of the spin chains}\label{speHS}

In this section we shall compute the partition function of the HS
spin chains of $BC_N$ type~\eqref{ssh} by using Polychronakos's
freezing trick~\cite{Po93,Po94} applied to the spin dynamical
models discussed in the previous sections. We shall first provide
a detailed heuristic justification of the freezing trick in the
present context. Our calculation relies on the computation of the
large coupling constant limit of the partition functions of the
spin dynamical models~\eqref{Hstar} and the scalar Sutherland
model of $BC_N$ type
\begin{equation}
\label{Hscalar}
\begin{aligned}
\Hs=-\sum_i \pa_{x_i}^2 &+ a(a-1)\,\sum_{i\neq
j}\left(\sin^{-2} x_{ij}^-+
\sin^{-2} x_{ij}^+\right)\\
&{}+b(b-1)\,\sum_i \sin^{-2}\!x_i+b'(b'-1)\,\sum_i \cos^{-2}\!x_i\
\end{aligned}
\end{equation}
acting on the Hilbert space $L^2_0(\tC)$.
Using the definition~\eqref{bebep} of $\be$ and $\be'$
we obtain
\begin{equation}\label{scalarH2}
\begin{aligned}
\Hs=-\sum_i \pa_{x_i}^2 &+ a^2\, U(\bx)-a\,V(\bx)\,,
\end{aligned}
\end{equation}
with $U(\bx)$ and $V(\bx)$ respectively given by Eqs.~\eqref{U} and~\eqref{V}.
{}From Eqs.~\eqref{sH2} and~\eqref{scalarH2} it follows that
\begin{equation}
\label{starH2}
H^*_{\ep\ep'}=-\sum_i\pa_{x_i}^2-a\,\ssH^*_{\ep\ep'} + a^2 U(\bx)
=\Hs+a\big(V(\bx)-\ssH^*_{\ep\ep'}\big)\,,
\end{equation}
where $H^*_{\ep\ep'}$ is assumed to act in the Hilbert space
$\cH=L^2_0(\tC)\otimes\cS$. Let $\bigl\{\psi_i(\bx)\bigr\}_{i\in\NN}$
be a basis of eigenfunctions of $\Hs$, and let
$\bigl\{\ket{\si_{\ep\ep',j}}\bigr\}_{j=1}^d$, with $d={(2M+1)}^N$, be
a basis of eigenfunctions of $\ssh_{\ep\ep'}$, so that
$$
\Hs\,\psi_i(\bx)=E_i\,\psi_i(\bx)\,,\qquad
\ssh_{\ep\ep'}\,\ket{\si_{\ep\ep',j}}=\sse_{\ep\ep',j}\,\ket{\si_{\ep\ep',j}}\,.
$$
The set
$\bigl\{\psi_i(\bx)\ket{\si_{\ep\ep',j}}\bigr\}_{i\in\NN,\;1\leq
j\leq d}$ is thus a basis of the Hilbert space $\cH$, and
$$
\Hs\bigl(\psi_i(\bx)\ket{\si_{\ep\ep',j}}\bigr)=
E_i\,\psi_i(\bx)\ket{\si_{\ep\ep',j}}\,,
$$
since $\Hs$ does not act on the spin variables.

{}From Appendix~\ref{AppB} it follows that the classical potential
$a^2\,U(\bx)-a\,V(\bx)$ has a unique equilibrium point (actually, a minimum)
$\bchi(a)$ in the set $\tC$ provided that
$a>\max(1/\be,1/\be',1)$. The freezing trick is based on the fact
that for $a\gg1$ the eigenfunctions of $\Hs$ are all sharply
peaked around the equilibrium $\bchi(a)$. Since
$\bchi(a)=\bxi+O\big(a^{-1}\big)$, for $a\gg1$ we have
\begin{multline*}
  \bigl[V(\bx)-\ssH^*_{\ep\ep'}\bigr]
  \bigl(\psi_i(\bx)\ket{\si_{\ep\ep',j}}\bigr)
  =\psi_i(\bx)\bigl[V(\bx)-\ssH^*_{\ep\ep'}\bigr]\ket{\si_{\ep\ep',j}}\\[1mm]
  \simeq
  \psi_i(\bx)\bigl[V(\bxi)-\ssH^{*0}_{\ep\ep'}\bigr]\ket{\si_{\ep\ep',j}}
  =\psi_i(\bx)\big(\ssh_{\ep\ep'}\ket{\si_{\ep\ep',j}}\big)
  =\sse_{\ep\ep',j}\,\psi_i(\bx)\ket{\si_{\ep\ep',j}}\,.
\end{multline*}
Thus for $a\gg 1$ the Hamiltonian $H^*_{\ep\ep'}$ is approximately diagonal in
the basis $\bigl\{\psi_i(\bx)\ket{\si_{\ep\ep',j}}\bigr\}_{i\in\NN,\;1\leq j\leq
    d}\,$, with eigenvalues approximately given by
\begin{equation}\label{Eij}
E_{\ep\ep',ij}^*\simeq E_i+a\,\sse_{\ep\ep'\!,j}\,,
\qquad i\in\NN\,,\quad 1\leq j\leq d\,,\quad a\gg1\,.
\end{equation}
Taking into account that $\sse_{\ep\ep',j}$ is independent of $a$, we
immediately obtain the following \emph{exact} expression for the eigenvalues
$\sse_{\ep\ep',j}$ of the spin chain Hamiltonian~\eqref{ssh}:
\begin{equation}\label{sse}
\sse_{\ep\ep'\!,j}=\lim_{a\to\infty}\frac1a\,(E^*_{\ep\ep'\!,ij}-E_i)\,.
\end{equation}
Using Eq.~\eqref{Eij}, we can easily derive the ground state energy
$\sse_{\ep\ep'\!,\text{min}}$ of the spin chains~\eqref{ssh}.
Indeed, clearly the ground state energy $E^*_{\ep\ep'\!,\text{min}}$
given by Eq.~\eqref{Emin} is achieved when both $E_i$ and $\sse_{\ep\ep'\!,j}$
in Eq.~\eqref{Eij} attain their minimum values $E_{\text{min}}$
and $\sse_{\ep\ep'\!,\text{min}}$. {}From Eqs.~\eqref{Enstar} and~\eqref{bebep}
it follows that both $E_{\ep\ep',ij}^*$ and $E_i$ are polynomials
of the second degree in $a$ with the same leading coefficient.
Hence
\begin{equation}\label{ssemin}
\sse_{\ep\ep'\!,\text{min}}=\lim_{a\to\infty}\frac1a\,(E^*_{\ep\ep'\!,\text{min}}
-E_{\text{min}})\,.
\end{equation}
Since $E_{\text{min}}$ is obtained by setting $n=0$ in Eq.~\eqref{Enstar},
the coefficient of $a$ in $E_{\text{min}}$ vanishes. {}From the previous equation,
it follows that $\sse_{\ep\ep'\!,\text{min}}$ is the coefficient of $a$ in
$E^*_{\ep\ep'\!,\text{min}}$, namely (cf.~Eq.~\eqref{Emin})
\begin{multline}
\sse_{\ep\ep'\!,\text{min}}=m_0\big[
4N^2+4(2\Bbe-1)N+\frac\kappa3\,\big(\kappa-2m_0(6\Bbe+2m_0\kappa-3)\big)\\
-2\rho(2\Bbe+m_0\kappa-1)-\rho^2 \big],\label{eminlong}
\end{multline}
where $\kappa$, $\rho$ and $m_0$ are defined in Eqs.~\eqref{murho}
and~\eqref{mzero}, and we have set
\begin{equation}
  \label{bbar}
  \Bbe =\frac12(\be+\be')\,.
\end{equation}
Note that (as remarked in the previous section) the ferromagnetic
ground state energy vanishes, since $m_0=0$ when $\ep=1$.

We emphasize that Eq.~\eqref{sse} cannot be used directly to compute in full
the spectrum of $\ssh_{\ep\ep'}$, since it is not clear {\em a priori} which
eigenvalues of $H^*_{\ep\ep'}$ and $H_s$ can be combined to yield an
eigenvalue of $\ssh_{\ep\ep'}$. The importance of Eq.~\eqref{sse} lies on the
fact that it can be used as the starting point for the exact computation of
the partition function of the spin chain $\ssh_{\ep\ep'}$, which in turn
completely determines the spectrum.

Let us denote by $\Zs$, $Z^*_{\ep\ep'}$ and $\ssZ_{\ep\ep'}$ the
partition functions of the scalar Sutherland Hamiltonian
\eqref{Hscalar}, the Sutherland spin dynamical model
\eqref{Hstar}, and the spin chain Hamiltonian \eqref{ssh},
respectively. {}From Eq.~\eqref{Eij} it follows that
$Z^*_{\ep\ep'} (T)\simeq\Zs(T)\,\ssZ_{\ep\ep'} (T/a)$, and hence
\begin{equation}
\label{ssZ}
\ssZ_{\ep\ep'} (T)=\lim_{a\to\infty}\frac{Z^*_{\ep\ep'}(aT)}{\Zs(aT)}\,.
\end{equation}
Recall that the spectrum of the Hamiltonian $H^*_{\ep\ep'}$ of the
spin dynamical model is given by Eq.~\eqref{Enstar}, where
$n\in[\ZZ^N]$ is a nonnegative nonincreasing multiindex
(satisfying conditions \eqref{cond1} if $\ep=-1$), and
$s=(s_1,\dots,s_N)$, $-M\leq s_i\leq M$, satisfies
\eqref{cond2}--\eqref{cond3}.  The leading terms in the expansion
of $E^*_{\ep\ep'}(n;s)$ are therefore
\begin{equation}
  \label{Enapp}
  E^*_{\ep\ep'}(n;s)\simeq a^2E_0+8a\sum_in_i(\Bbe+N-i)\,,
\end{equation}
where $E_0=4\sum_i(\Bbe+N-i)^2$ is a constant independent of $n$. The
eigenvalues $E(n)$ of the scalar Sutherland Hamiltonian $\Hs$ are given by the
right-hand side of Eq.~\eqref{Enstar}, where the multiindex $n\in[\ZZ^N]$ is
now unrestricted. Thus $E(n)$ also satisfies Eq.~\eqref{Enapp} for $a\gg1$.

Let us start by computing the large $a$ limit of the denominator in
Eq.~\eqref{ssZ}.  Note, first of all, that the constant $E_0$ can be dropped
from both $E^*_{\ep\ep'}(n;s)$ and $E(n)$ without affecting the value of
$\ssZ_{\ep\ep'}(T)$.  Using the asymptotic expansion of $E(n)$ and setting
\begin{equation}
  \label{qdef}
  q = \e^{-8/(k_{\mathrm B}T)}
\end{equation}
we immediately obtain
\begin{equation}
  \label{ZsaT}
  \Zs(aT)\simeq \sum_{n\in[\ZZ^N]}\,q^{\sum\limits_i n_i(\Bbe+N-i)}\,.
\end{equation}
Defining $p_i=n_i-n_{i+1}$, $1\leq i\leq N-1$, and $p_N=n_N$ we
have
$$
\prod_i q^{n_i(\Bbe+N-i)}=\prod_{i\leq j}q^{p_j(\Bbe+N-i)}=\prod_j
q^{p_j\sum\limits_{i=1}^j(\Bbe+N-i)}=\prod_j
q^{j p_j\big(\Bbe+N-\tfrac12(j+1)\big)}
$$
and hence
\begin{align}
\label{ZsaTfinal}
\Zs(aT)&\simeq
\sum_{p_1,\dots,p_N\geq0}\,\prod_i q^{i p_i\big(\Bbe+N-\tfrac12(i+1)\big)}
=\prod_i \sum_{p_i\geq0} q^{i p_i\big(\Bbe+N-\tfrac12(i+1)\big)}\notag\\
&=\prod_i\bigg[1-q^{i\big(\Bbe+N-\tfrac12(i+1)\big)}\bigg]^{-1}.
\end{align}

Let us compute next the partition function $Z^*_{\ep\ep'}(aT)$ of
the Sutherland spin dynamical model \eqref{Hstar} for $a\gg1$. To
this end, it is convenient to represent the multiindex
$n\in[\ZZ^N]$ appearing in Eq.~\eqref{Enapp} as
\begin{equation}
\label{nm}
n =
\big(\overbrace{\vphantom{1}m_1,\dots,m_1}^{k_1},
\overbrace{\vphantom{1}m_2,\dots,m_2}^{k_2},\dots,
\overbrace{\vphantom{1}m_r,\dots,m_r}^{k_r}\big),
\end{equation}
where $m_1>m_2>\cdots>m_r\geq0$, and $k_i=\#(m_i)\in\NN$ satisfies
$k_1+\cdots+k_r=N$ (together with condition \eqref{cond1}, if $\ep=-1$). Thus
\begin{align*}
  E^*_{\ep\ep'}(n;s)&\simeq 8a\sum_{i=1}^rm_i\sum_{j=
    k_1+\cdots+k_{i-1}+1}^{k_1+\cdots+k_{i-1}+k_i}(\Bbe+N-j)\\
  &=8a\sum_{i=1}^r m_i k_i\bigg(\Bbe+N-\frac12-\frac{k_i}2-\sum_{j=1}^{i-1}k_j\bigg)
  \equiv8a\sum_{i=1}^r m_i\nu_i\,.
\end{align*}
Let $k=(k_1,\dots,k_r)$, and denote by $d_{\ep\ep'}(k,m_r)$ the cardinal of
the set of spin quantum numbers $s$ satisfying conditions
\eqref{cond2}--\eqref{cond3} for the multiindex \eqref{nm}, namely
\begin{subequations}
\label{dk}
\begin{align}
\label{dkp}
d_{\ep\ep'}(k,m_r)&=\textstyle
\binom{2M+1+\de_{1,\ep}(k_1-1)}{k_1}\cdots
\binom{2M+1+\de_{1,\ep}(k_r-1)}{k_r}\,,\qquad m_r>0\,;\\[2mm]
\label{dk0}
d_{\ep\ep'}(k,0)&=\textstyle
\binom{2M+1+\de_{1,\ep}(k_1-1)}{k_1}\cdots
\binom{2M+1+\de_{1,\ep}(k_{r-1}-1)}{k_{r-1}}
\binom{M_{\ep'}+\de_{1,\ep}(k_r-1)}{k_r}\,.
\end{align}
\end{subequations}
The partition function $Z^*_{\ep\ep'}(aT)$ is therefore given by
\begin{align*}
  &Z^*_{\ep\ep'}(aT)\simeq\sum_{k\in\cP_N}
  \sum_{m_1>\cdots>m_r\geq0} d_{\ep\ep'}(k,m_r)
  \prod_{i=1}^r q^{m_i\nu_i}\\
  &{}=\sum_{k\in\cP_N}
  \sum_{m_1>\cdots>m_r>0}d_{\ep\ep'}(k,m_r)\prod_{i=1}^r q^{m_i\nu_i}
  +\sum_{k\in\cP_N}
  \sum_{m_1>\cdots>m_{r-1}>0} d_{\ep\ep'}(k,0)
  \prod_{i=1}^{r-1} q^{m_i\nu_i},
\end{align*}
where we have denoted by $\cP_N$ the set of partitions of the positive integer
$N$. Since
\begin{align*}
\sum_{m_1>\cdots>m_s>0}\prod_{i=1}^s q^{m_i\nu_i}
&=\sum_{p_1,\dots,p_s>0}\prod_{i=1}^s q^{\nu_i \sum\limits_{j=i}^s p_j}
=\sum_{p_1,\dots,p_s>0}\prod_{i=1}^s\prod_{j=i}^s q^{p_j\nu_i}\\
&=\sum_{p_1,\dots,p_s>0}\prod_{j=1}^s q^{p_j\sum\limits_{i=1}^j\nu_i}
=\prod_{j=1}^s\sum\limits_{p_j>0}q^{p_j\sum\limits_{i=1}^j\nu_i}
={\ds\prod\limits_{j=1}^s\frac{q^{N_j}}{1-q^{N_j}}}\,,
\end{align*}
where
\begin{equation}\label{Nj}
N_j=\sum_{i=1}^j\nu_i=\Big(\sum_{i=1}^jk_i\Big)\Big(\Bbe
+N-\frac12-\frac12\sum_{i=1}^jk_i\Big)\,,
\end{equation}
using Eqs.~\eqref{dk} we finally obtain
\begin{multline}\label{Z*aTfinal}
Z^*_{\ep\ep'}(aT)\simeq\sum_{(k_1,\ldots,k_r)\in\cP_N}\bigg\{
\bigg[{\textstyle\binom{M_{\ep'}+\de_{1,\ep}(k_r-1)}{k_r}
+\binom{2M+1+\de_{1,\ep}(k_r-1)}{k_r}}
\,\frac{q^{N_r}}{1-q^{N_r}}\bigg]\\
\times\prod_{j=1}^{r-1}\bigg[{\textstyle\binom{2M+1+\de_{1,\ep}(k_j-1)}{k_j}}\,\frac{q^{N_j}}{1-q^{N_j}}\bigg]\bigg\}\,.
\end{multline}
Equations~\eqref{ssZ}, \eqref{ZsaTfinal} and~\eqref{Z*aTfinal}
yield the following exact
formula for the partition function of the HS spin
chain~\eqref{ssh}:
\begin{align}
  \ssZ_{\ep\ep'}(T)&=\prod_{i=1}^N
  \bigg[1-q^{i\big(\Bbe+N-\tfrac12(i+1)\big)}\bigg]
  \sum_{(k_1,\ldots,k_r)\in\cP_N}\bigg\{
  \bigg[\textstyle\binom{M_{\ep'}+\de_{1,\ep}(k_r-1)}{k_r}\notag\\
  &\quad{}+{\textstyle\binom{2M+1+\de_{1,\ep}(k_r-1)}{k_r}}
  \,\frac{q^{N_r}}{1-q^{N_r}}\bigg]
  \prod_{j=1}^{r-1}\bigg[{\textstyle\binom{2M+1+\de_{1,\ep}(k_j-1)}{k_j}}\,\frac{q^{N_j}}{1-q^{N_j}}\bigg]\bigg\}\,.
\label{Zchainfinal}
\end{align}

{}From the previous formula, which is in fact the main result of
this paper, one can infer several remarkable properties of the
spectrum of the spin chain~\eqref{ssh} that we shall now discuss.
First of all, for half-integer $M$ the partition
function~\eqref{Zchainfinal} does not depend on $\ep'$, since in
this case $M_\pm=M+1/2$. Hence the spectrum of the spin
chain~\eqref{ssh} is independent of $\ep'$ when $M$ is a
half-integer, a property that is not immediately apparent from the
expression of the Hamiltonian~\eqref{ssh}. Secondly, all the
denominators $1-q^{N_k}$, $1\leq k\leq r$, appearing in the second
line of Eq.~\eqref{Zchainfinal} are included as factors in the
product in the first line. Hence the partition
function~\eqref{Zchainfinal} can be rewritten as
\begin{equation}\label{Zsimple}
\ssZ_{\ep\ep'}(T)=\sum_{\de\in\{0,1\}^N}d_{\ep\ep'\!,\de}(M)\,q^{\bvep_\de}\,,
\end{equation}
where $\bvep_\de$ is given by
\begin{equation}\label{possspec}
\bvep_\de=\sum_{i=1}^N\de_i\,i\big(\Bbe+N-\frac12\,(i+1)\big)\,,\qquad
\de=(\de_1,\ldots,\de_N)\,,
\end{equation}
and the degeneracy factor $d_{\ep\ep'\!,\de}(M)$ is a polynomial
of degree $N$ in $M$. Therefore, for {\it all} values of $\ep$,
$\ep'$ and $M$, the spectrum of the spin chain is
contained in the set of $2^N$ numbers $8\bvep_\de$,
$\de\in\{0,1\}^N$. Moreover, for generic (sufficiently large)
values of the spin $M$, the spectrum \emph{exactly} coincides with
the above set of numbers, the values of $\ep$, $\ep'$ and $M$
affecting only the degeneracy $d_{\ep\ep'\!,\de}(M)$ of each
level. In particular, from the previous observation and
Eqs.~\eqref{hahf} and~\eqref{possspec} we immediately obtain the
following exact expression for the constant $V(\bxi)$ (i.e.,~half
the maximum energy of the antiferromagnetic chain):
\begin{equation}\label{Vofxi}
V(\bxi)=4\sum_{i=1}^N i\big(\Bbe+N-\frac12\,(i+1)\big)
=\frac23\,N(N+1)(2N+3\Bbe-2)\,.
\end{equation}
{}From Eq.~\eqref{possspec} it follows that the energies of the spin
chains~\eqref{ssh} are of the form $8(j\,\Bbe+k)$, with $j,k$ nonnegative
integers.  Since the coefficients of the powers of $q$ appearing in
Eq.~\eqref{Z*aTfinal} are independent of $\Bbe$, it follows that for
generic\!\footnote{More precisely, for all real values of $\Bbe$ except for a
  finite (possibly empty) set of rationals.} values of $\Bbe$ the degeneracy
of the levels depends only on the spin $M$ and the number of particles $N$.

\section{Discussion and conjectures}\label{dis-conj}

In this section we shall present several concrete examples in
which we apply the formula~\eqref{Zchainfinal} to compute the
spectrum of the spin chains~\eqref{ssh} for certain values of $N$
and $M$. These examples strongly suggest a number of conjectures
that shall be discussed in detail at the end of this section. We
shall restrict ourselves to the antiferromagnetic chains
$\ssh_{-,\pm}$, the properties of their ferromagnetic counterparts
following easily from the relation~\eqref{hahf}.

{\ni\itshape\bfseries Example 1.\;} The structure of Eq.~\eqref{Zchainfinal} makes
it straightforward to compute the spectrum of the spin chains for
any~{\it fixed} number of particles as a function of the spin.
For instance, for $N=3$ sites and integer $M$ the energies
(divided by $8$) of the spin chain $\ssh_{--}$ are
$0,\,\Bbe+2,\,2\Bbe+3,\,3\Bbe+3,\,3\Bbe+5,\,4\Bbe+5,\,
5\Bbe+6,\,6\Bbe+8$, with respective degeneracies
\begin{align*}
&\tfrac16\,M(M-1)(M-2),\;\tfrac56\,M(M^2-1),\;\tfrac16\,M(M+1)(11M-2),\\[1mm]
&\tfrac16\,M(M+1)(7M-4),\;\tfrac16\,M(M+1)(7M+11),\;\tfrac16\,M(M+1)(11M+13),\\[1mm]
&\tfrac56\,M(M+1)(M+2),\;\tfrac16\,(M+1)(M+2)(M+3).
\end{align*}
Note that in this case all energy levels $8\bvep_\de$, with
$\bvep_\de$ given by~\eqref{possspec}, are attained for $M\geq 3$,
in agreement with the general discussion of the previous section.
For $M=1$ and a few values of $\be$ and $\be'$, we have
numerically computed the spectrum of the spin chain~\eqref{ssh} by
representing the operators $S_{ij}$ and $S_i$ as $27\times 27$
matrices. The results obtained are in complete agreement with those listed
above.

For a fixed value of the spin $M$, we have not been able to find
an explicit formula expressing the energies and their degeneracies
as functions of the number of particles $N$. However, if we fix
$M$ the spectrum can be straightforwardly computed from
Eq.~\eqref{Zchainfinal} for any given value of $N$. We shall
next present two concrete examples for the cases $M=1/2$ and
$M=1$.

\begin{figure}[h]
\psfrag{e}{$\sse_i$}
\psfrag{d}{$d_i$}
\includegraphics[width=11cm]{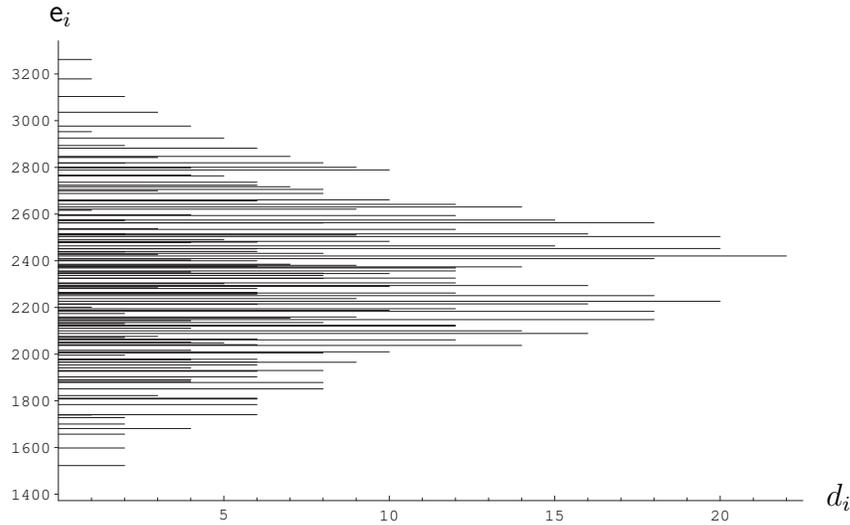}
\begin{quote}
\caption{Energy levels $\sse_i$ and degeneracies $d_i$
of the antiferromagnetic spin $1/2$ chain $\ssh_{-,\pm}$
for $N=10$ particles and $\Bbe=\sqrt 2$.\label{spin12N10ed}}
\end{quote}
\end{figure}

{\ni\itshape\bfseries Example 2: spin \mathversion{bold}$1/2$.\;}
In this case we have computed the partition function
$\ssZ_{-,\pm}$ for up to 20 particles (recall that for
half-integer $M$ the chains with $\ep'=\pm1$ have the same
spectrum). For instance, for $N=6$ the antiferromagnetic spin
chain energies (divided by $8$) and their corresponding
degeneracies (denoted by subindices) are given by
\begin{align*}
&(9\Bbe+32)_2,\:
(10\Bbe+36)_2,\:
(11\Bbe+38)_2,\:
(11\Bbe+41)_4,\:
(12\Bbe+38)_1,\:
(12\Bbe+43)_6,\\
&(13\Bbe+43)_3,\:
(13\Bbe+46)_6,\:
(14\Bbe+46)_4,\:
(14\Bbe+50)_4,\:
(15\Bbe+47)_2,\:
(15\Bbe+50)_3,\\
& (15\Bbe+55)_6,\:
(16\Bbe+51)_2,\:
(16\Bbe+55)_5,\:
(17\Bbe+53)_1,\:
(17\Bbe+56)_4,\:
(18\Bbe+58)_3,\\
& (19\Bbe+61)_2,\:
(20\Bbe+65)_1,\:
(21\Bbe+70)_1.
\end{align*}
The number of levels increases rapidly with the number of
particles $N$. For example, if $N=10$ the number of levels (for
generic values of $\Bbe$) is $136$, while for $N=20$ this number
becomes $7756$. It is therefore convenient to plot the energy
levels $\sse_i$ and their corresponding degeneracies $d_i$, as is
done in Fig.~\ref{spin12N10ed} for $N=10$ particles. Note that
Eq.~\eqref{possspec} implies that when $\Bbe\gg N$ the levels
cluster around integer multiples of $8\Bbe$. In fact, for all $N$
up to $20$ we have observed that these integers take \emph{all}
values in a certain range $j_0,j_0+1,\ldots,N(N+1)/2$; for
example, in the case $N=6$ presented above $j_0=9$.

\smallskip

\begin{figure}[h]
\psfrag{e}{$\sse_i$}
\psfrag{d}{$d_i$}
\includegraphics[width=11cm]{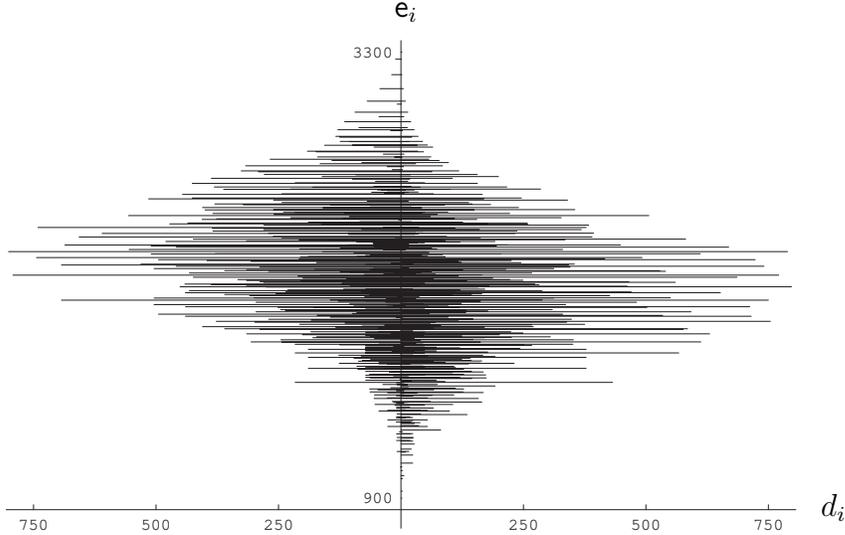}
\begin{quote}
\caption{Comparison of the energy levels $\sse_i$ and degeneracies
$d_i$ of the antiferromagnetic spin $1$ chains $\ssh_{--}$ (left) and $\ssh_{-+}$ (right) for
$N=10$ particles and $\Bbe=\sqrt 2$.\label{spin1N10ed}}
\end{quote}
\end{figure}

{\ni\itshape\bfseries Example 3: spin \mathversion{bold}$1$.\;}
We have computed the partition functions $\ssZ_{-,\pm}$ of the
spin chains $\ssh_{-,\pm}$ with spin $M=1$ for up to $15$
particles. As remarked in the previous section, for integer $M$
the partition functions $\ssZ_{-,\pm}$ are expected to be
essentially different.  This is immediately apparent from
Fig.~\ref{spin1N10ed}, where we have compared
graphically the energy spectra of the even and odd spin chains
$\ssZ_{-,\pm}$ with $\Bbe=\sqrt 2$ for $N=10$ particles.
However, we shall prove in what follows that the standard deviation of the
energy is \emph{exactly} the same for both chains. This rather unexpected
result will be relevant in the ensuing discussion of the level
density (see Conjecture~2 below). We also note that,
just as for spin $1/2$, for $N$ up to (at least) $15$ and \mbox{$\Bbe\gg N$}
the energy levels cluster around an
equally spaced set of nonnegative integer multiples of $8\Bbe$.

The previous examples for spin $1/2$ and $1$ suggest several
conjectures on the spectrum of the (antiferromagnetic) HS
spin chains of $BC_N$ type that we shall now present and discuss in detail.

{\ni\itshape\bfseries Conjecture 1.\;} \textsl{For $\Bbe\gg N$, the
energies cluster around an equally spaced set of levels of the form
$8j\ms\Bbe$, with $j=j_0,j_0+1,\ldots,N(N+1)/2$}.

In fact, for sufficiently large values of the spin $M$ this
assertion (with $j_0=0$) follows directly from
Eq.~\eqref{possspec}. Our calculations for a wide range of values
of $N$ and $M$ fully corroborate the above conjecture.

{\ni\itshape\bfseries Conjecture 2.\;} \textsl{For $N\gg 1$,
the level density follows a Gaussian distribution}.

\begin{figure}[h]
\psfrag{F}{$F_{\cN}(\sse),\,F(\sse)$}
\psfrag{e}{$\sse$}
\includegraphics[width=11cm]{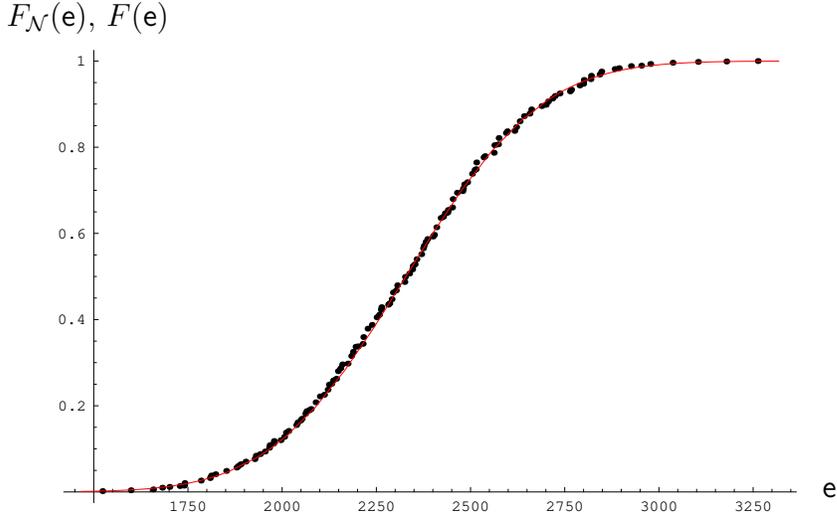}
\begin{quote}
\caption{Distribution functions $F_{\cN}(\sse)$ (continuous line)
and $F(\sse)$ (at its discontinuity points) for $\Bbe=\sqrt 2$, $M=1/2$, and
$N=10$.\label{normalspin12N10}}
\end{quote}
\end{figure}

More precisely, the number of levels (counting their
degeneracies) in an interval $I$ is approximately given
by
\begin{equation}\label{Numlevels}
(2M+1)^N \int_I \cN(\sse;\mu,\si)\d\sse\,,
\end{equation}
where
\begin{equation}\label{gaussian}
\cN(\sse;\mu,\si)=\frac1{\si\sqrt{2\pi}}\,\e^{-\frac{(\sse-\mu)^2}{2\si^2}}
\end{equation}
is the normal (Gaussian) distribution with parameters $\mu$ and
$\sigma$ respectively equal to the mean and standard deviation of
the energy spectrum of the spin chain. Although the shape of the plots in
Figs.~\ref{spin12N10ed} and~\ref{spin1N10ed} make this conjecture quite plausible,
for its precise verification it is preferable to compare the distribution function
\begin{equation}\label{FcN}
F_{\cN}(\sse)=\int_{-\infty}^\sse \cN(t;\mu,\si)\d t
\end{equation}
of the Gaussian probability density with its discrete analogue
\begin{equation}\label{F}
F(\sse)=(2M+1)^{-N}\sum_{i;\,\sse_i\leq\sse} d_i\,,
\end{equation}
where $d_i$ denotes the degeneracy of the energy level $\sse_i$.
Indeed, our computations for a wide range of values of $M$ and
$N\gtrsim 10$ are in total agreement with the latter conjecture for all four chains~\eqref{ssh}.
This is apparent, for instance,
in the case $\Bbe=\sqrt 2$, $M=1/2$, and $N=10$ presented in Fig.~\ref{normalspin12N10}.
The agreement between the distribution functions~\eqref{FcN} and~\eqref{F}
improves dramatically as $N$ increases. In fact, their plots are
virtually undistinguishable for $N\gtrsim 15$.

It is well known in this respect that a Gaussian level density is a characteristic feature of the
``embedded Gaussian ensemble'' (EGOE) in Random Matrix
Theory~\cite{MF75}. It should be noted, however, that the EGOE
applies to a system of $N$ particles with up to $n$-body
interactions ($n<N$) in the high dilution regime $N\to\infty$,
$\ka\to\infty$ and $N/\ka\to 0$, where $\ka$ is the number of
one-particle states. Since in our case $\ka=2M+1$ is fixed, the
fact that the level density is Gaussian does not follow from the
above general result. A study of the energy spectrum of the spin
chains~\eqref{ssh} in the framework of Random Matrix Theory is
nonetheless worth undertaking, and will be the subject of a
subsequent publication.

If Conjecture~2 is true, the level density for large $N$ is completely characterized
by the parameters $\mu$ and $\si$ through the Gaussian
law~\eqref{gaussian}. It is therefore of great interest to compute these
parameters in closed form as functions of $N$ and $M$. To this end,
let us write
\begin{equation}\label{hmm}
\ssh_{-,\pm}=\sum_{i\neq
j}\Big[h_{ij}(1+S_{ij})+\tih_{ij}(1+\tS_{ij})\Big] +\sum_i
h_i(1\mp S_i)\,,
\end{equation}
where the constants $h_{ij}$, $\tih_{ij}$ and $h_i$ can be easily
read off from Eq.~\eqref{ssh}. We shall begin by computing the
average energy $\mu_{--}\equiv\mu_-$ of the odd antiferromagnetic
spin chain~$\ssh_{--}$ for integer spin. Using the formulas for
the traces of the spin operators given in Table~\ref{traces} we
immediately obtain
\begin{align}
\mu_-&=(2M+1)^{-N}\tr\ssh_{--}
=\frac{2(M+1)}{2M+1}\,\Big[\sum_{i\neq j}(h_{ij}+\tih_{ij})+\sum_i h_i\Big]\notag\\
&=\frac{2(M+1)}{2M+1}\,V(\bxi)
=\frac{4(M+1)}{3(2M+1)}\,N(N+1)(2N+3\Bbe-2)\,,\qquad
M\in\NN\,,\label{mum}
\end{align}
where we have used the explicit expression~\eqref{Vofxi} for $V(\bxi)$.
\begin{table}[h]
\caption{Traces of products of the spin
operators.\vskip2mm}\label{traces}
\begin{tabular}{lll}\hline
  \vrule height 15pt depth 9pt width0pt Operator & Trace (integer $M$)
  & Trace (half-integer $M$)\\ \hline
 $S_i\ms$ & $(2M+1)^{N-1}$ & 0\\ \hline
$S_{ij}\ms$,\, $\tS_{ij}$ & $(2M+1)^{N-1}$ & $(2M+1)^{N-1}$\\
\hline
$S_iS_j$ & $(2M+1)^{N-2+2\de_{ij}}$ & $(2M+1)^N\de_{ij}$\\ \hline
$S_{ij}S_k\ms$,\, $\tS_{ij}S_k\ms$ & $(2M+1)^{N-2}$ & 0\\ \hline
$S_{ij}\tS_{kl}$ & $(2M+1)^{N-2}$ &
$(2M+1)^{N-2}(1-\de_{ik}\de_{jl})(1-\de_{il}\de_{jk})$\\ \hline
$S_{ij}S_{kl}\ms$,\, $\tS_{ij}\tS_{kl}$ &
$(2M+1)^{N-2+2\de_{ik}\de_{jl}+2\de_{il}\de_{jk}}$ &
$(2M+1)^{N-2+2\de_{ik}\de_{jl}+2\de_{il}\de_{jk}}$\\ \hline
\end{tabular}
\end{table}
On the other hand, the average energy $\mu_{-+}\equiv\mu_+$ of the even chain $\ssh_{-+}$
is given by
\begin{align}
\mu_+&=(2M+1)^{-N}\tr\ssh_{-+}
=\frac{2(M+1)}{2M+1}\,\sum_{i\neq j}(h_{ij}+\tih_{ij})
+\frac{2M}{2M+1}\,\sum_i h_i\notag\\
&=\frac2{2M+1}\,\big[(M+1)V(\bxi)-\Si_1\big]\,,\qquad M\in\NN\,,
\label{mup}
\end{align}
where $\Si_1\equiv\sum_i h_i$ is obviously independent of the
spin $M$. Similarly, for half-integer spin the formulas for the traces of the
spin operators in Table~\ref{traces} yield
the following expression for the mean energy
$\mu_{-,\pm}\equiv\mu_{\pm}$:
\begin{equation}\label{mupmhalf}
\mu_{\pm}=\frac1{2M+1}\,\big[2(M+1)V(\bxi)-\Si_1\big]\,,\qquad
M=\frac12\,,\,\frac32\,,\,\dots
\end{equation}

Let us turn now to the (squared) standard deviation of the energy, given by
\[
\si_{-,\pm}^2\equiv\si_\pm^2=\frac{\tr(\ssh_{-,\pm}^2)}{(2M+1)^N}
-\frac{(\tr\ssh_{-,\pm})^2}{(2M+1)^{2N}}\,.
\]
For integer spin, a long but straightforward calculation using
the formulas in Table~\ref{traces} yields
\begin{equation}
\si_\pm^2=\frac{4M(M+1)}{(2M+1)^2}\,\Big[2\sum_{i\neq
j}\big(h_{ij}^2+\tih_{ij}^2\big)+\sum_ih_i^2\Big]\equiv
\frac{4M(M+1)}{(2M+1)^2}\,\Si_2\,,\qquad M\in\NN\,.\label{sigmapm}
\end{equation}
Since $\Si_2$ does not depend on $M$, the above equation
completely determines the dependence of $\si_\pm$ on the spin.
An important consequence of the previous formula
is the equality of the standard deviation of the energy for the even and odd
antiferromagnetic chains (for half-integer spin, this follows trivially
from the fact that the even and odd chains have the same spectrum).
This result is quite surprising, since for integer spin the energy spectra of
the chains $\ssh_{-,\pm}$ are essentially different, cf.~Fig.~\ref{spin1N10ed}.
For half-integer spin, an analogous calculation yields the expression
\begin{equation}\label{sipmhalf}
\si_\pm^2=\frac{4M(M+1)}{(2M+1)^2}\,\Big[\Si_2+\frac{\Si_3}{M(M+1)}\Big]\,,
\qquad M=\frac12\,,\,\frac32\,,\,\dots,
\end{equation}
where
\begin{equation}
\Si_3=\frac14\,\sum_i h_i^2-\sum_{i\neq j}h_{ij}\tih_{ij}
\end{equation}
is independent of the spin. As before, Eq.~\eqref{sipmhalf} fixes the
dependence of $\si_\pm$ on the spin.

We still need to evaluate $\Si_1(N)$, $\Si_2$(N) and $\Si_3(N)$ in order to determine
the dependence on $N$ of $\mu_\pm$ and $\si_\pm$ in all cases.
Although we have not been able to
compute these quantities in closed form, in view of Eq.~\eqref{mum}
it is natural to formulate the following conjecture:

{\ni\itshape\bfseries Conjecture 3.\;} \textsl{The average energy $\mu_\pm$
and its squared standard deviation $\si_\pm^2$ depend polynomially on $N$.}

In fact, since $\sse_{-,\pm\ms;\ms\text{max}}=2V(\bxi)$
is a polynomial of degree $3$ in $N$ by Eq.~\eqref{Vofxi},
it follows that the degrees in $N$ of $\mu_\pm$ and $\si_\pm^2$
cannot exceed $3$ and $6$, respectively. The latter conjecture and this fact
allow us to determine the quantities $\Si_i(N)$ by evaluating $\mu_\pm$ and $\si_\pm^2$
for $N=2,\dots,8$ and $M=1/2,1$ using the exact formula~\eqref{Zchainfinal} for the
partition function (cf.~Eqs.~\eqref{mup}, \eqref{sigmapm} and~\eqref{sipmhalf}).
The final result is
\begin{align}
\Si_1&=\Si_3=2N(2\Bbe+N-1)\,,\notag\\
\Si_2&=\frac{4N}9\,\big[\,2(2N^2+3N+13)\ms\Bbe^{\,2}+
(N-1)(5N^2+7N+20)\ms\Bbe\label{Sigmas}\\
&\hphantom{{}=\frac{4N}9\,\big[\,2(2N^2+3N}%
{}+\frac15\,(N-1)(8N^3+3N^2+13N-12)\ms\big]\,.\notag
\end{align}
These expressions, together with Eqs.~\eqref{mum}--\eqref{sipmhalf},
completely determine $\mu_\pm$ and $\si_\pm$ for \emph{all} values of $M$ and $N$.
We have checked that the resulting formulas yield the exact values
of $\mu_\pm$ and $\si_\pm$ computed from the partition function~\eqref{Zchainfinal}
for a wide range of values of $M$ and $N$. This provides a very solid confirmation
of Conjecture~3. Let us mention, in closing,
that formulas analogous to \eqref{mum}--\eqref{sipmhalf} expressing the mean and standard
deviations of the energy for the ferromagnetic chains
$\ssh_{+,\pm}$ can be immediately deduced from the previous
expressions and Eq.~\eqref{hahf}.

\begin{ack}
This work was partially supported by the DGI under grant
no.~BFM2002--02646. A.E. acknowledges the financial support of the
Spanish Ministry of Education and Science through an FPU
scholarship. The authors would like to thank J.~Retamosa for
pointing out Ref.~\cite{MF75} to them.
\end{ack}

\appendix
\section{Uniqueness of the equilibrium of the classical potential}\label{AppB}

In this Appendix we shall prove that the classical potential~\eqref{U}
has exactly one equilibrium point in the set $\tC$. We have already
seen in Section~\ref{intHS} that $U$ has at least one minimum in
$\tC$. We shall now prove that the Hessian of $U$ is positive-definite
in $\tC$, which implies that all the critical points of $U$ in $\tC$
must be minima. This implies that $U$ has exactly one critical point
(a minimum) in the set $\tC$.

If $f(t)=\sin^{-2}t$, we can express the second partial derivatives of $U$
as follows
\begin{equation}
\begin{aligned}
\frac{\pa^2 U}{\pa x_i^2}&=2\sum_{j\neq i}\,\big[
f''(x_{ij}^-)+f''(x_{ij}^+)\big]+\be^2 f''(x_i)+{\be'}^2 f''\big(\frac\pi 2-x_i\big)\,,\\
\frac{\pa^2 U}{\pa x_i\pa x_j}&=2\,\big[f''(x_{ij}^+)-f''(x_{ij}^-)\big]\,.
\end{aligned}
\end{equation}
Note that $f''(t)=2\csc^2 t(1+3\cot^2 t)$ is strictly positive for
all values of $t$, and therefore $(\pa^2 U)/(\pa x_i^2)>0$ for all
$i$. By Gerschgorin's theorem~\cite[15.814]{GR00}, the eigenvalues
of the Hessian of $U$ lie in the union of the intervals
\[
\Big[\,\frac{\pa^2 U}{\pa x_i^2}-\ga_i,\frac{\pa^2 U}{\pa x_i^2}+\ga_i\,\Big]\,,
\quad\text{where}\quad \ga_i=\sum_{j\neq i}\Big|\,\frac{\pa^2 U}{\pa x_i\pa x_j}\,\Big|\,,\qquad
i=1,\ldots,N.
\]
Since
\[
\frac{\pa^2 U}{\pa x_i^2}-\ga_i\geq\be^2 f''(x_i)+{\be'}^2 f''\big(\frac\pi 2-x_i\big)>0,
\]
all the eigenvalues of the Hessian of $U$ are strictly positive.
This establishes our claim.

\end{document}